%% file: main.tex
\documentclass[letterpaper,twocolumn,10pt]{article}
\usepackage{usenix,url,amsfonts,amsmath,amsthm,multirow,amssymb,bmpsize}
\usepackage{graphicx}
\usepackage{subcaption}
\usepackage{enumitem}
\usepackage{booktabs}
\usepackage{listings}  
\usepackage{url}

\usepackage{breakurl}
\usepackage[breaklinks]{hyperref}
\usepackage{xcolor}
\setlist{noitemsep}
\graphicspath{{figs/}}
\DeclareGraphicsExtensions{.png}

\usepackage{adjustbox}
\usepackage{array}
\usepackage{float}
\usepackage[margin={1.5cm,1.5cm}]{geometry}
\usepackage[leftmargin=2em]{quoting}

\newcommand\mytt[1]{\texttt{\small{#1}}}

\newif\ifblind
\blindfalse

\newif\ifccs
\ccsfalse

\newcolumntype{R}[2]{%
    >{\adjustbox{angle=#1,lap=\width-(#2)}\bgroup}%
    l%
    <{\egroup}%
}

\begin{document}

\date{}

\title{Towards more accurate and useful data anonymity vulnerability measures}

\author{
Paul Francis$^{\dag}$ \quad David Wagner$^{\S}$ \\
$^{\dag}$Max Planck Institute for Software Systems (MPI-SWS), Germany\\
$^{\S}$Deutsche Universit{\"a}t f{\"u}r Verwaltungswissenschaften\\
francis@mpi-sws.org, dwagner@uni-speyer.de
}

\maketitle

\input{abstract}
\input{introduction}
\input{precision}
\input{allowed-inference}
\input{baseline}
\input{base-rate}

\input{discussion}
\input{summary}

\bibliographystyle{abbrv}
\bibliography{../../masterBib/master}

\end{document}

%% file: abstract.tex
\begin{abstract}
  The purpose of anonymizing structured data is to protect the privacy of individuals in the data while retaining the statistical properties of the data. There is a large body of work that examines anonymization vulnerabilities. Focusing on strong anonymization mechanisms, this paper examines a number of prominent attack papers and finds several problems, all of which lead to overstating risk. First, some papers fail to establish a correct statistical inference baseline (or any at all), leading to incorrect measures. Notably, the reconstruction attack from the US Census Bureau that led to a redesign of its disclosure method made this mistake. We propose the non-member framework, an improved method for how to compute a more accurate inference baseline, and give examples of its operation.
  
  Second, some papers don't use a realistic membership base rate, leading to incorrect precision measures if precision is reported. Third, some papers unnecessarily report measures in such a way that it is difficult or impossible to assess risk.  Virtually the entire literature on membership inference attacks, dozens of papers, make one or both of these errors. We propose that membership inference papers report precision/recall values using a representative range of base rates.
  
\end{abstract}

%% file: introduction.tex
\section{Introduction}
\label{sec:introduction}

An important and heavily-researched problem is that of anonymizing structured data. The goal is to preserve the statistical properties of the data while protecting the anonymity of individual persons in the data.

The literature consists of mechanisms to anonymize data and attacks on both those mechanisms and real deployments. There are hundreds of papers of both types. The attack papers primarily focus on describing \textit{vulnerabilities}: weaknesses in the design that could potentially be exploited. Attack papers also frequently comment informally on \textit{risk}: the likelihood that a vulnerability will in fact be exploited in the wild.

It is important that attack papers measure vulnerability correctly. If vulnerability is underestimated, then anonymized data may be reidentified by bad actors. If vulnerability is overestimated, then the utility of data may be reduced, either because data is too-aggressively anonymized or because it is not released at all.

It is also important that attack papers present vulnerability measures in such a way that risk can be easily assessed. Attack papers often suggest that their attacks lead to serious risks, and it is important that readers, including non-experts, can judge for themselves whether this is the case for their own scenario.

This paper examines a number of attacks and finds instances where either vulnerability is not measured correctly, or measures are presented in such a way that risk cannot be evaluated. Regarding incorrect measures, we identify two problems, \textit{allowed inference errors} and \textit{base rate errors}.

An allowed inference error occurs when an attack measure fails to recognize an inference as an intended statistical inference. For example, suppose an attack achieves 80\% precision on predictions of sex given that job title is `Computer Scientist' simply by always predicting `male'. This can hardly be regarded as a loss of anonymity. Indeed, one purpose of the anonymous data release might well be to know the distribution of sex across different jobs. Measures of attack effectiveness need to at least show that the attack is better than an allowed statistical guess. The US Census Bureau's decision to redesign its disclosure methodology in 2020 is the result of an allowed inference error. Section~\ref{sec:allowed-inference} gives several examples of these errors.

To address this problem, we propose the \textit{non-member framework}, a novel approach to computing a \textit{baseline allowed inference}. This is the baseline inference quality that must be exceeded for an attack to be regarded as a vulnerability. The non-member framework is based on the well-established principle that a dataset does not leak privacy for individuals who are not in the dataset (and are independent of others in the dataset). The idea is to find the best inferences that can be made from a dataset about individuals not in the dataset. Because these inferences do not leak privacy about the (non-member) individuals, the resulting precision serves as a privacy-neutral baseline. Section \ref{sec:baseline} sketches out the non-member framework and presents a simple demonstration.

A base rate error is when an unrealistic base rate is used to report precision. For example, consider a membership attack which has a false positive rate of 0.05 (5\% of non-members are incorrectly predicted to be members), and a false negative rate of 0.0. If half of the population of 2000 individuals being tested are members, then we get 1000 true positives (TP), and 50 false positives (FP), and $precision = TP/(TP+FP) = 1000/1050 = 0.952$. But if only 1\% of the tested population are members, then we get $TP=20$ and $FP=99$, and the precision drops to around 0.17. Virtually all membership attack papers use an unrealistic base rate. This produces an incorrect precision when precision is reported.

Many membership attack papers report ROC rather than precision. ROC can in principal be converted to precision/recall, but the vast majority of membership attack papers do not present data that would allow assessment of high-precision/low-recall attacks. Problems with membership attack papers are covered in Section~\ref{sec:base-rate}.

We argue that data anonymization vulnerability papers should always report precision and recall or recall-like measures. Doing so both facilitates risk assessment and allows for apples-to-apples comparisons between different attacks.

Altogether, this paper makes the following contributions:
\begin{itemize}
    \item The non-member framework for establishing baseline allowed inferences, which is more accurate, more general, and more efficient than former approaches.
    \item A description of how two inference attacks in the literature exhibit the allowed inference error (and a third which is not new).
    \item An argument that data anonymity attack papers can and should report precision and recall-like measures.
    \item Identification of the base rate error in membership inference attack papers, and a demonstration of using precision and recall to improve reporting.
\end{itemize}


%% file: precision.tex
\section{Precision and coverage are general measures of vulnerability}
\label{sec:precision}

All things being equal, an attack paper that facilitates risk assessment is better than one that does not. As part of a risk assessment, it is useful to know:
\begin{enumerate}
    \item How is an attack carried out?
    \item What attributes about individuals in a dataset can be learned (predicted) by an attack?
    \item How likely is it that an attack's predictions are correct?
    \item For what fraction of individuals can said predictions be made?
\end{enumerate}

This information helps a stakeholder reason about an attacker's cost/benefit; how likely it is that an attack might be launched, and what damage results. While it isn't the role of a vulnerability study to assess risk per se, it is certainly helpful if the measures presented in an attack paper answer these questions directly. Of particular interest for the purpose of this paper are the third and fourth items. We are interested in two types of predictions; \textit{inference predictions}, where an attacker predicts a target individual's unknown, or \textit{secret}, attribute, and \textit{membership predictions}, where an attacker predicts a target's membership in a dataset.

In what follows, we describe various options for measuring the third and fourth items: the likelihood that a prediction is correct, and the fraction of individuals for which predictions can be made.

The third item could be measured as \textit{accuracy A}, defined as true (correct) predictions over all predictions:
\begin{equation}
    accuracy \ A \ = \ true \ predictions / \ all \ predictions
\label{eq:accuracy}
\end{equation}
Accuracy normally measures both positive and negative predictions. In measuring privacy, however, the predictions of concern are normally positive predictions; the attacker is interested in what the target's value is, not what it is not. We therefore prefer the measure of \textit{precision P} over accuracy, defined as:
\begin{equation}
    precision \ P \ = \ true \ PP / \ all \ PP
\label{eq:precision}
\end{equation}
where $PP$ is the number of positive predictions.

In the case of predicting a value from a continuous distribution, we can define a prediction as true if it is within a certain range, or within a certain amount of error.

The fourth item can be measured as \textit{recall R} or a similar measure. In the context of binary decisions, for instance membership predictions, recall is defined as:
\begin{equation}
    recall \ R \ = \ TP \ / \ TP\!+\!FN \ = \ TP/all \ positives
    \label{eq:recall}
\end{equation}
where $TP$ is the number of true positive predictions and $FN$ the number of false negative predictions.

Inference predictions, however, are not usually binary. One way around this is to compute recall separately for each attribute value. For instance, say the attribute is education, and the choices are elementary, high-school, college, masters, and phd. To calculate recall for the value college, the data would be cast as binary values college and not-college, and any predictions other than college are treated as negative predictions.

We generally prefer, however, a measure which we call \textit{prediction rate PR} which can be used instead of recall. Here, we allow the attacker to make \textit{no prediction} rather than a negative prediction. Prediction rate $PR$ is defined as:
\begin{align}
    prediction \ rate \ PR &= TP \ / \ TP\!+\!FP\!+\!np \\
                         & = TP/all \ possible \ predictions
    \label{eq:prediction-rate}
\end{align}
where $np$ represents the choice to make \textit{no prediction} where a prediction was otherwise possible.

An example is the following. Suppose that a given attack only works on outliers of some sort (say individuals with very high salaries). Given a non-outlier target, the attacker chooses not to make any prediction at all, and this non-prediction reduces the prediction rate. In other words, the fact that there are very few attackable records means that the vulnerability is not widely applicable, and this is reflected in a lower prediction rate measure.

Given that there are multiple ways to compute recall-like measures, in the paper we refer to all such measures informally as \textit{coverage C}. Precision alone or coverage alone is an incomplete measure. A high precision attack may still constitute very little risk if the coverage is extremely small, especially if the coverage is random (versus specific targets such as high-income individuals).


Throughout this paper, we demonstrate how precision and coverage serve as general-purpose measures for data anonymity vulnerabilities.

%% file: allowed-inference.tex
\section{Allowed Inference Errors}
\label{sec:allowed-inference}

This section describes three different attacks that mistakenly claim that anonymity was violated because of allowed inference errors.

\subsection{US Census reconstruction attack}
\label{sec:census-attack}

The highest profile allowed inference error is the US Census Bureau attack that triggered the redesign of the 2020 census.

\begin{figure}
\begin{center}
\includegraphics[width=1.0\linewidth]{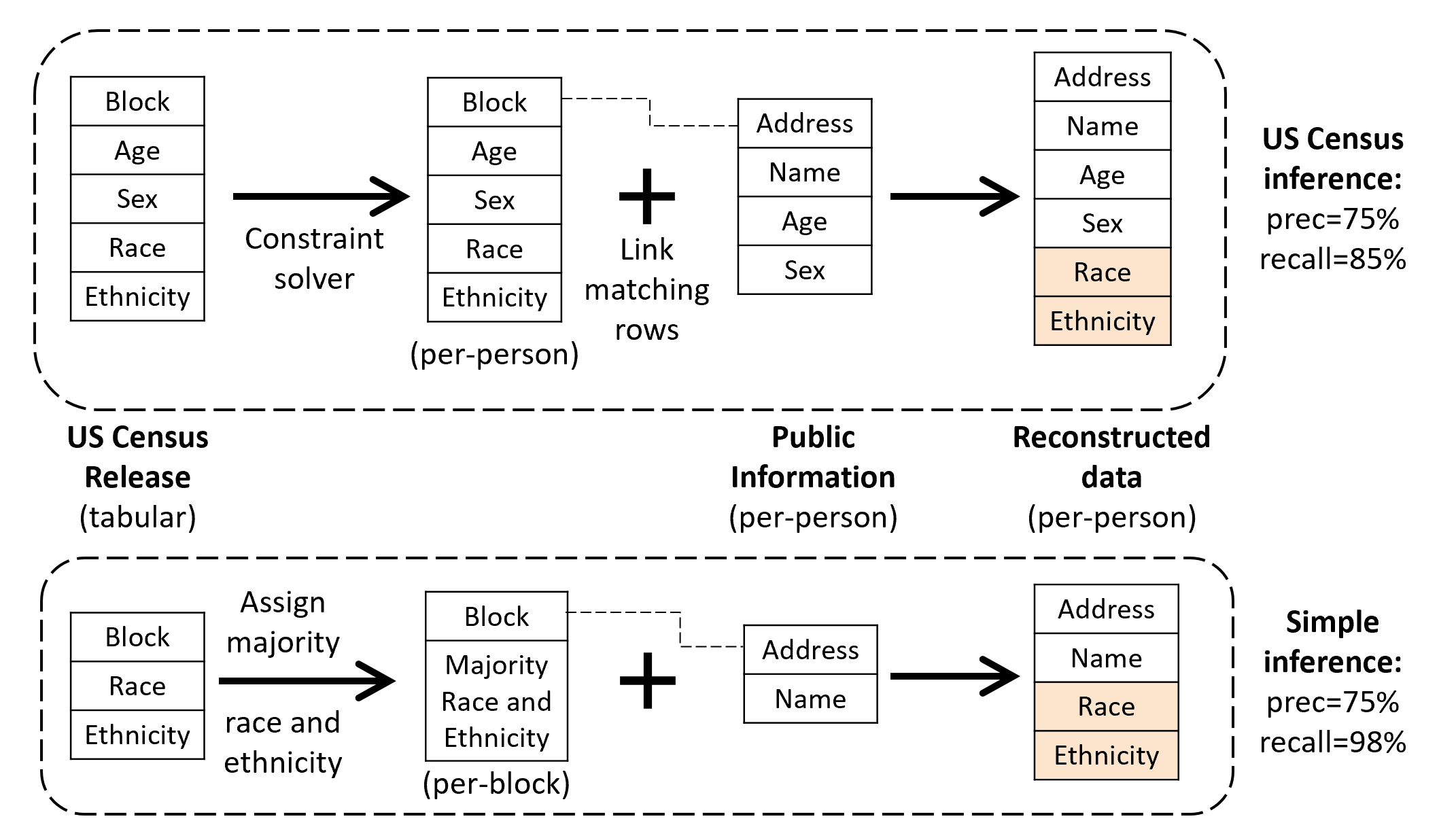}
\caption{The US Census attack compared to simple inference from the per-block majority race and ethnicity~\cite{francis2022census}.
}
\label{fig:census-attack}
\end{center}
\end{figure}

The attack is illustrated in Figure~\ref{fig:census-attack}. The Bureau releases tabular data (attributes plus counts) per census block. Each block covers a geographical area whose population can range from a few individuals to hundreds of individuals.

The Bureau's attack takes as input the block, age, sex, race, and ethnicity as published by the Bureau as tabular data. It runs a constraint solver over this data to produce a record-level (per-person) reconstruction of the same attributes. It links this with publicly available data including name, address (which maps to block), age, and sex. This linkage allows it to associate race and ethnicity with an identified person.

The results of this attack were initially published using a \emph{reconstruction measure}. This is a measure of the fraction of reconstructed records for which there are matching original records. In 2019, Abowd\footnote{Abowd was CTO at the US Census Bureau at the time.} reported that 17\% of reconstructed records were exact matches, assuming publicly available data for linking~\cite{abowd-2019-staring}.

One problem with reporting only a reconstruction measure is that it is not necessarily clear what the precision and recall are. Does 17\% reconstruction mean a precision of 17\% against 100\% recall, a precision of 100\% against 17\% recall, or something in between? The nature of the vulnerability cannot be understood without this information.

More generally, reconstruction percentage alone is not a reliable measure of anonymity, and should not be used in isolation. Consider for instance the outcome of an election: 500 votes for candidate A, and 600 votes for candidate B. The votes could be perfectly reconstructed simply by writing down 500 A's and 600 B's. Nevertheless the election results are anonymous. Reconstruction percentage is an accurate measure of vulnerability when every prediction is for a set of attributes unique to each individual, as was the case with Dinur et al.~\cite{dinur2003revealing}. Some confusion might have been avoided, however, if Dinur et al. had used precision instead of reconstruction percentage.

In 2021, in response to a lawsuit from the state of Alabama against the US Census Bureau, Abowd finally published precision and recall results~\cite{lawsuit-alabama}. For publicly available linkage data, the attack achieved $P=0.38$ and $R=0.18$, and for perfect linkage data (i.e. the Census' internal data), $P=0.75$ and $R=0.64$.

Although described as a reconstruction attack, this is essentially an inference attack, where the known attributes are address, age, and sex, and the inferred secret attributes are race and ethnicity.

The Bureau regarded this as an unacceptable level of inference, and so redesigned its anonymization mechanism for the 2020 census. The prior mechanism relied primarily on aggregation and swapping, whereas the new mechanism relies primarily on aggregation and noise and produces a differential privacy measure.

In 2022, Francis showed that \textit{better} precision/inference can be had by merely assigning each block's majority race and ethnicity to every person in the block~\cite{francis2022census}. For this inference, the known attribute is address (from which block is derived); age and sex are not even required.

This simple inference has a precision/recall of $P=0.75$ and $R=0.98$, assuming perfect address information. This performance is possible for the simple reason that census blocks tend to be homogeneous with respect to race and ethnicity, and so predicting the majority race and ethnicity for all individuals is a good guess. Indeed by limiting the measure to only those blocks with a single race/ethnicity, the simple inference achieves $P=1.0$ at $R=0.11$. 

Francis showed that this simple inference works equally well on the new differential privacy mechanism as the old swapping mechanism, as should be the case since differential privacy aims to retain the statistical properties of the data.

Francis argued that, because the Bureau considers it acceptable to release majority race/ethnicity statistics about blocks, inferring race/ethnicity using majority statistics is acceptable and should not be regarded as a privacy violation. Francis conjectured that it is possible that the entire success of the Bureau's reconstruction attack could be due to correct inferences on majority race/ethnicity individuals. He could not test this conjecture, however, without access to the US Census data.

In response to Francis' and others' criticisms of the Bureau's measurement methodology~\cite{ruggles2021role,muralidhar2022reexamination}, Abowd et al. implemented a new reconstruction attack~\cite{abowd2023census}. Here, Abowd et al. now agree with Francis that inferences on block-majority race/ethnicity individuals should not be regarded as privacy violations. As such, Abowd et al. report detail about block-minority race/ethnicity subpopulations.

The new attack was able to obtain high precision ($P=0.95$) on a specific subpopulation; individuals within blocks where the solver had only one solution and who were unique in the block with respect to sex and age-bin (see Table 10B in~\cite{abowd2023census}). For this subpopulation, the attack obtained a recall of $R=0.003$ (908K individuals) for publicly available linkage data, and $R=0.01$ (3.3M individuals) for perfect linkage data.

While the new attack is far less effective than the original attack claimed, this precision is nevertheless well above a simple statistical baseline, and so represents a meaningful vulnerability so far as we know (though arguably not a meaningful risk).

Note that as of this writing, no evidence has been presented that the original reconstruction attack constituted a meaningful vulnerability. We don't know how the US Census Bureau would have responded had it correctly understood the vulnerability from the beginning.

\subsection{Inferring location traces}
\label{sec:mobility}

Mobile telecommunications companies routinely sell anonymized mobility data\footnote{A web search for ``purchase geolocation data'' produces many such data sets.}. The raw source data is typically the location of the base station where each phone connected, timestamp, and phone ID. The data is typically anonymized by aggregating timestamps and locations into larger units (e.g. one hour windows and one km square grid), and releasing counts of the number of phones in the resulting aggregates, usually with low counts suppressed.

In 2017, Xu et al. reported that this method of anonymization is not private~\cite{xu2017trajectory}. From the abstract:
\begin{noindent}
\begin{quoting}
\textit{the attack system is able to recover users’ trajectories with accuracy about 73\%$\sim$91\% at the scale of tens of thousands to hundreds of thousands users, which indicates severe privacy leakage in such dataset.}
\end{quoting}
\end{noindent}

A trajectory (or trace) is defined as the sequence of time-window/ location-square aggregates, where the location assigned for a given window is the one where the device spent the most time. Note that the trace includes every time window, regardless of whether the location changed or not.

Per-trace accuracy is measured as the fraction of time-window/ location-squares in a reconstructed trace that have a match in the linked original trace. Accuracy is the average of these. Original and reconstructed traces are linked by selecting the best match in a greedy fashion.

Xu et al. reconstruct traces by predicting the next location using simple assumptions about the general characteristics of human mobility. These characteristics are 1) at night people stay in the same location, 2) in the day, the next location is best predicted by the prior location, direction, and velocity, and 3) locations on a given day are similar to that of the previous day. Although these characteristics are validated on the original data itself, the authors argue that they are general in nature and so could be used for other datasets.

What made us suspect that the attack measures might not exceed a baseline is that the general characteristics of human mobility are not specific to any one person. As a thought experiment, imagine that one person was removed from the original dataset before the mobility characteristics were validated. The resulting validation would not be significantly different from the validation with that person included. We should therefore not expect any reconstruction based only on general mobility characteristics to accurately capture individual deviations from those characteristics.

The paper goes on to say:
\begin{noindent}
\begin{quoting}
\textit{given the two most frequent locations of the recovered trajectories, over 95\% of them can be uniquely distinguished. Therefore, the results indicate that the recovered trajectories are very unique and vulnerable to be reidentified with little external information.}
\end{quoting}
\end{noindent}

According to Xu et al., the two most frequent locations are home and work: these are therefore the known attributes. The secret attributes are any other locations in the trace. Xu et al. implies that an attacker could on average learn 73\% of a target's secret locations knowing only the target's home and work locations for 95\% of all targets.

The mistake here is that the measure doesn't adjust for what is already known---the target's home and work locations. If we assume\footnote{We could not obtain the original data to validate this assumption, nor could Xu et al. validate it for us.} that on average people spend 8 hours at work and 10 hours at home, then this already constitutes 75\% of the target's trace (18 hours / 24 hours = 0.75).

What is likely happening is this: Most traces are dominated by home and work location. Therefore, when linking original and reconstructed traces, naturally those with the same home and work locations are linked. These two locations can easily account for 73\% of the trace accuracy, which suggests that the locations that are neither home nor work are \textit{not} accurately reconstructed.

At a minimum, Xu et al. should have published a reconstruction measure that excludes what is assumed to be known. It would have been better still if they had published precision measures for specific predictions and compared the results with statistical guesses (i.e. if the attacker knows the home location, what is the precision of predicting the work location).

As the paper stands, it provides no evidence that publishing aggregate counts of mobility data leads to privacy-violating trace inference.

\subsection{Inference attacks exploiting ML models}
\label{sec:warfarin}

ML models are used to determine the dosage of the drug warfarin given the clinical history, demographics, and certain genetic markers of a patient. Fredrikson et al. suggest that this model can be used in reverse to learn patient's genetic markers, and therefore is a privacy risk~\cite{ristenpart-pharmacogenetics}. The basic idea is that, given patient demographics like age, race, height and weight, and the warfarin dosage, the model can be used to predict the patient's genetic markers.

A key result is that, using the paper's model reversing technique, an attacker is able to improve the accuracy of predictions of a given genetic marker by 22\% over what can be predicted using only marginal distributions of the same data, which is publicly available information often published in studies~\cite{ristenpart-pharmacogenetics}.

Unlike the prior two examples, which didn't consider an allowed inference baseline at all, Fredrikson et al. are using these marginal distributions as a baseline.  If there is another allowed inference that makes better predictions, however, then that should be used as the baseline instead.  It so happens that Fredrikson et al. measured just such another baseline, this one based on running the attack against members of the validation set instead of members of the training set. The attack is only 3\% more accurate against this non-training baseline.

The reason this non-training baseline is an allowed baseline stems from a widely accepted principle of data privacy that a dataset that does not include a given individual should not be regarded as leaking privacy about that individual (see \S\ref{sec:non-member}). Since the non-training individuals are not among the data used for the model, the predictions on non-training individuals leak no privacy and therefore can serve as an allowed baseline.

The 3\% attack accuracy improvement on members of the model's training set over the allowed baseline is negligible, and so this attack should not be regarded as a vulnerability, much less a privacy risk.

%% file: baseline.tex
\section{Computing the allowed inference baseline}
\label{sec:baseline}

This section describes the basic concept of the non-member framework for computing an allowed inference baseline, and gives an example of its usage using the BankChurners dataset from Kaggle\footnote{https://www.kaggle.com/code/amanpatyal/exploratory-analysis-bankchurners-csv}. A more thorough exploration of the non-member framework is left for future work.

Section~\ref{sec:allowed-inference} described three allowed inference errors. In the case of the US Census (\S\ref{sec:census-attack}), we presume that it is the Census Bureau's intent to release the majority race and ethnicity for each block. In other words, our criteria for determining the allowed baseline is the Bureau's policy.

While in this case the policy is self-evidently correct, in general policy is not an appropriate criteria for determining an allowed baseline. A stakeholder cannot arbitrarily decide that it is ok to release data that leads to high precision/coverage inferences, and then declare by fiat that such inferences are not vulnerabilities.

Fredrikson et al. used data published in other studies (marginal distributions) as an implicit allowed baseline (\S\ref{sec:warfarin}). Though not stated explicitly, the implication is that this data is a legitimate allowed baseline \textit{because} it is published elsewhere without apparent problems. We take it as self-evident, however, that the mere existence of data published elsewhere does not mean that it is allowed. After all, perhaps that data is itself not anonymous.

The above two examples suggest that 1) the allowed baseline must be computed from the dataset being anonymized, and 2) there must be some technical criteria for establishing allowed baselines (not just some policy).

\subsection{Core concept behind the non-member framework}
\label{sec:non-member}

The non-member framework is based on the following core concept:
\begin{noindent}
\begin{quoting}
\textbf{If an individual is not present in a dataset, and is independent of all other individuals in the dataset, then the release of that dataset does not violate that individual's privacy.}
\end{quoting}
\end{noindent}

This is not to say that no harm can come to a dataset non-member from the release of the dataset. A dataset that says that smokers have higher health care costs can lead to an increase in a smoker's health insurance premiums even if the smoker is not part of the dataset. Nevertheless, the dataset does not violate the individual's privacy per se.

This concept is similar to the core concept of differential privacy, which is
that a individual's presence or absence in an anonymized dataset should not substantially alter that individual's privacy risk~\cite{dwork2017exposed}:
\begin{noindent}
\begin{quoting}
\textit{Our ultimate privacy goal when releasing information about a sensitive dataset is to ensure that anything that can be learned about an individual from the released information can be learned without that individual's data being included.}
\end{quoting}
\end{noindent}

If we accept the core concept, then it follows that any inference made about non-members resulting from an analysis of the dataset is an allowed inference, and can serve as a statistical baseline: inferences about dataset members made by an attack on the anonymized dataset must be better than inferences made about non-members from the original data. If they are not, then the attack cannot be regarded as a vulnerability.

The need for dataset non-member independence from dataset members is important. If member Mary and non-member Nate have identical data, then identifying Mary in the dataset allows an attacker to make correct inferences about Nate. A baseline precision based on these inferences would be too high. As a practical matter, however, it appears possible to mitigate the problem (see \S \ref{sec:dependence}).

\subsection{Basic framework}

\begin{figure}
\begin{center}
\includegraphics[width=0.9\linewidth]{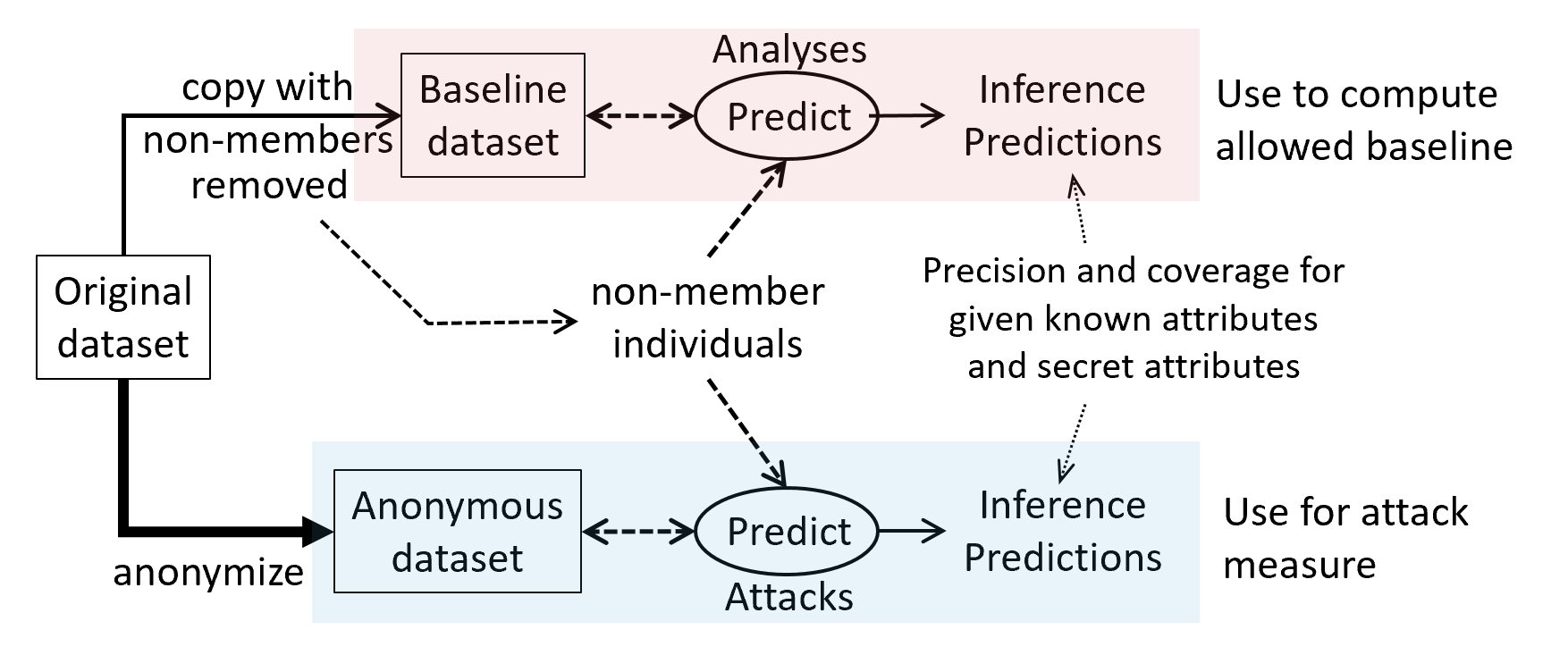}
\caption{The non-member framework for computing allowed baselines. For any given combination of known attributes and secret attributes, the analysis that produces the best precision and coverage is used as the allowed baseline.
}
\label{fig:non-member}
\end{center}
\end{figure}

Figure~\ref{fig:non-member} illustrates our approach to computing the baseline. One or more \textit{non-member} individuals are selected from an \textit{original dataset}. A \textit{baseline dataset} is created as a copy of the original dataset but with these non-members removed. For each non-member, using known attributes of the non-member, an \textit{analysis} is run over the baseline dataset to \textit{predict} the non-member's secret attributes.

These predictions define a set of allowed baseline precision and coverage pairs $P_{base}^{\mathbb{A}_i}$ and $C_{base}^{\mathbb{A}_i}$, where $\mathbb{A}_i$ are the $i^{th}$ conditions of the analysis including at least the known attributes and the secret attributes. The conditions can also be more specific, including for instance specific values of the secret or known attributes.

So long as the non-members are independent from the members, any analyses are legitimate. The analyses that produce the best inference predictions serve as the best baseline. Since there could always be better analyses, any computed baseline precision and coverage is a lower bound on the baseline. If there might be some dependence between non-members and members, then the analysis should avoid overfitting (\S \ref{sec:dependence}).

A natural approach, an example of which is given in \S\ref{sec:non-member-example}, is to fit an ML model as the analysis. A complete version of this would be to remove one individual at a time, fitting the ML model and predicting the individual's secret each time. A more efficient approach would be to remove multiple individuals, fit the ML model, and then do the predictions on that model. This latter relaxed approach should generally produce a baseline close enough to that of the complete approach, at a significant savings in computation.

\subsection{Comparing the baseline to an attack}

Separately, the original dataset (i.e. containing the non-members) can be anonymized to produce an \textit{anonymous dataset}. Attacks may be executed on the anonymous dataset to predict the non-members' secrets given known attributes. These predictions define an attack precision and coverage pairs $P_{atk}^{\mathbb{A}_i}$ and $C_{atk}^{\mathbb{A}_i}$. The attack precision and coverage must exceed the baseline precision and coverage for an attack to be considered a vulnerability.

Strictly speaking, the baseline analysis is independent of the anonymous dataset and associated attacks in the sense that neither are directly used in the baseline analysis. To be able to compare the attack and baseline scores, however, it is necessary that the conditions $\mathbb{A}_i$ match. Because of this, any given analysis will normally be done in the context of a given attack and its conditions.

There are a variety of ways to compare the attack measures $P_{atk}^{\mathbb{A}_i}$ and $C_{atk}^{\mathbb{A}_i}$ with the base measures $P_{base}^{\mathbb{A}_i}$ and $C_{base}^{\mathbb{A}_i}$. A measure used by
\ifblind
\else
Francis et al.~\cite{francis2022diffix} is \textit{Precision Improvement PI}.
\fi
$PI$ compares the precision measures $P_{atk}^{\mathbb{A}_i}$ and $P_{base}^{\mathbb{A}_i}$ where $C_{atk}^{\mathbb{A}_i} = C_{base}^{\mathbb{A}_i}$:
\begin{equation}
    PI^{\mathbb{A}_i} \ = \ (P_{atk}^{\mathbb{A}_i} - P_{base}^{\mathbb{A}_i}) / (1 - P_{base}^{\mathbb{A}_i})
    \label{eq:precision-improvement}
\end{equation}

$PI$ measures the ratio of actual improvement to the best possible improvement. For example, both pairs [$P_{atk} = 0.75$; $P_{base} = 0.5$], and [$P_{atk} = 0.97$; $P_{base} = 0.94$] have $PI = 0.5$. $PI$ represents the loss of privacy caused by the anonymous dataset, independent of the absolute value of the attack precision. The three values $PI^{\mathbb{A}_i}$, $P_{atk}^{\mathbb{A}_i}$, and $C_{atk}^{\mathbb{A}_i}$ can be used by a stakeholder to assess risk.

For continuous variables, predictions are deemed correct when they are within a certain error $\epsilon$ of the true value. The comparison between $P_{base}^{\mathbb{A}_i}$ and $P_{atk}^{\mathbb{A}_i}$ must of course use the same $\epsilon$. In so doing, however, the resulting precision improvement measure is independent from the error chosen. This is important for geolocation data, where in one case any prediction worse than one kilometer might be considered safe, where in another case simply predicting the correct hemisphere might be a privacy violation.

To force $C_{atk}^{\mathbb{A}_i} = C_{base}^{\mathbb{A}_i}$ so that $PI^{\mathbb{A}_i}$ can be computed for a given attack, only the set of targets for which predictions are made in the attack are used to measure baseline precision. The procedure is as follows:
\begin{enumerate}
    \item A random set of targets $\mathbb{T}$ is selected from the original dataset (or alternatively all individuals in the original dataset if coverage is likely to be small).
    \item The attack is run for each target. Each individual attack may result in a prediction or in no prediction. The set of individuals for which a prediction was made are placed in set $\mathbb{P}$
    \item Define coverage $C^{\mathbb{A}_i}$ for both the attack and the baseline as the prediction rate $PR=|\mathbb{P}|/|\mathbb{T}|$
    \item Measure $P_{base}^{\mathbb{A}_i}$ using the targets in $\mathbb{P}$.
\end{enumerate}

\subsection{An example of the non-member framework}
\label{sec:non-member-example}

This section provides a simple example of the non-member framework. The purpose here is not to do a thorough exploration of the properties of inference baselines, but rather to illustrate the approach and give examples of various effects. A thorough exploration is needed, but saved for future work.
\footnote{All of the code and data for this section can be found at
\ifblind
\else
 \href{https://github.com/yoid2000/non-member-framework-paper-code}{https://github.com/yoid2000/non-member-framework-paper-code}
\fi
}

For this example, we use the BankChurners dataset from Kaggle (see Figure~\ref{fig:colChars}). BankChurners was selected because it has a good variety of categorical and continuous attributes, and several columns that can be regarded as PII (personally identifying information). This variety makes it a good choice for demonstrating various characteristics of the non-member framework.  It has 10127 rows. Each row contains information about one banking customer.


\begin{figure}
\begin{center}
\begin{small}
\begin{tabular}{rlll}
    \toprule
    Attribute & Type & PII & Distinct Values \\
    \midrule
        Attrition\_Flag & cat &  & 2 \\
        Avg\_Open\_To\_Buy & cont &  & 5156 \\
        Avg\_Utilization\_Ratio & cont &  & 945 \\
        Card\_Category & cat &  & 4 \\
        Contacts\_Count\_12\_mon & cont &  & 7 \\
        Credit\_Limit & cont &  & 4727 \\
        Customer\_Age & cont & X & 44 \\
        Dependent\_count & cont & X & 6 \\
        Education\_Level & cat & X & 7 \\
        Gender & cat & X & 2 \\
        Income\_Category & cat &  & 6 \\
        Marital\_Status & cat & X & 4 \\
        Months\_Inactive\_12\_mon & cont &  & 7 \\
        Months\_on\_book & cont &  & 44 \\
        Total\_Amt\_Chng\_Q4\_Q1 & cont &  & 1062 \\
        Total\_Ct\_Chng\_Q4\_Q1 & cont &  & 770 \\
        Total\_Relationship\_Count & cont &  & 6 \\
        Total\_Revolving\_Bal & cont &  & 1847 \\
        Total\_Trans\_Amt & cont &  & 4156 \\
        Total\_Trans\_Ct & cont &  & 124 \\
     \bottomrule
 \end{tabular}
 \end{small}
 \caption{There are 6 categorical and 14 continuous attributes in BankChurners. Five attributes are PII.}
 \label{fig:colChars}
 \end{center}
 \end{figure}

Our analysis uses simple ML models produced using \mytt{sklearn} \mytt{LogisticRegression} (penalty L1, C 0.01, solver saga), and \mytt{Lasso} (alpha 0.1) for categorical and continuous attributes respectively.  For categorical attributes, we computed a second precision by always predicting the most common attribute value. We use the better of the two predictions.  Predictions for continuous attributes are considered correct if they are within 5\% of the true value.  The resulting \textit{precision} is defined by Equation~\ref{eq:precision}.

We did not run a complete computation (i.e. where the ML analysis is run separately for each individual non-member). Rather, a set of 3039 randomly selected non-members were removed, and the analysis was run against the remaining dataset. This is analogous to training and test datasets in ML. There were 7088 members (e.g. training set).

Note that we made no special effort to optimize the ML models. The goal here is to demonstrate the non-member framework, not to study how to produce the best possible predictions.

\begin{figure}
\begin{center}
\includegraphics[width=0.85\linewidth]{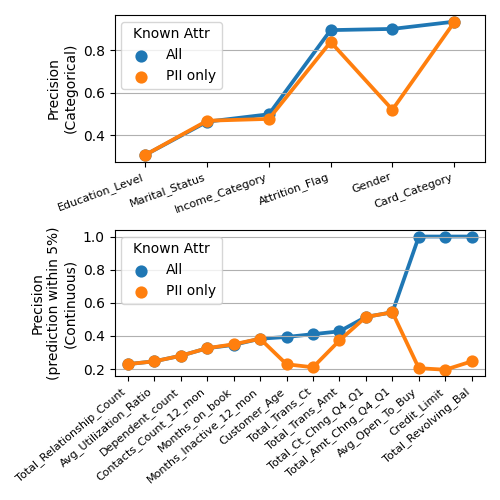}
\caption{Results for baseline precision achieved by simple ML predictions. The X axis is the target attribute. The features are either all attributes except the target, or only PII attributes.  Prediction rate $PR_{base} = 1.0$.
}
\label{fig:allVsPiiAcc}
\end{center}
\end{figure}

Figure~\ref{fig:allVsPiiAcc} shows the precision for both categorical and continuous secret attributes (prediction rate is 1.0).  Every attribute is used as the secret attribute in turn. The figure compares precision for two different sets of known attributes. One is where all attributes except the secret attribute are known. The other is where only the PII attributes are known. The latter represents the case where the attacker only has access to public information about the target, and the former represents worst-case attack knowledge.

There is nothing special about how these measures were computed: they represent a modest effort at ordinary ML modeling. They do, however, illustrate two points. First, given that some of the inference baselines are quite high, computing an inference baseline can make a huge difference in how an attack measure is interpreted. In these cases, a high-precision attack does not represent a severe vulnerability.  Second, it matters what attacker prior knowledge assumptions are made.

\subsection{Perfect precision}

None of the categorical attributes in Figure~\ref{fig:allVsPiiAcc} achieved perfect precision. We can, however, improve precision at the expense of a lower prediction rate, as follows. For categorical attributes, for each target record, the LogisticRegression predicts the probability for which each secret attribute value is the secret value. The predicted value is that with the highest probability $p_{max}$.

We could improve precision $P_{base}$ by defining a threshold $p_{thresh}$ whereby if $p_{max} < p_{thresh}$, then no prediction is made. This leads to a lower prediction rate $PR_{base}$ as defined in Equation~\ref{eq:prediction-rate}.

\begin{figure}
\begin{center}
\includegraphics[width=0.85\linewidth]{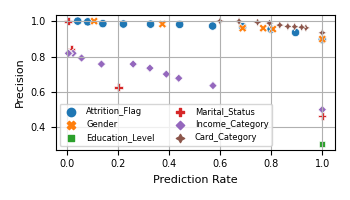}
\caption{Precision $P_{base}$ versus prediction rate $PR_{base}$ on the categorical variables of BankChurners. Each point for a given secret attribute represents a different cutoff threshold for making a prediction (versus making no prediction). Four of the six secret attributes achieve perfect precision.
}
\label{fig:prec-recall}
\end{center}
\end{figure}

Figure~\ref{fig:prec-recall} shows the result of manipulating precision and prediction rate using a variety of different thresholds $p_{thresh}$ for the categorical attributes of BankChurners as secrets. For 4 of the 6 categorical attributes, we could establish perfect precision.

The implication here is that even an attack with perfect precision $P_{atk}=1.0$ may not be a vulnerability, so long as there is a baseline with $P_{base}=1.0$ and equal or better coverage ($C_{base} \geq C_{atk}$) for the same individuals.

This might seem counter-intuitive, but in fact it should not be surprising. For example, it would not be surprising that a prediction of \mytt{degree='PhD'} given that \mytt{title='professor'} would have near-perfect precision, nor would high precision in this case be regarded as breaking anonymity.

\subsection{Mitigating the effect of dependent non-member and member individuals}
\label{sec:dependence}

As mentioned earlier, a direct dependence between non-member and member individuals can overestimate the baseline inference.

One way to mitigate this problem would be to search for individuals in the baseline dataset that are very similar to the removed non-member. If there are one or two that are substantially more similar than others, then these may be dependent individuals and so should also be removed.

A simpler and more scalable way to mitigate the problem, however, is to use an ML analysis that avoids overfitting, and therefore tends to hide the effect of dependent records.

To demonstrate this, we modified the original BankChurners dataset to exhibit varying degrees of dependence. We built three datasets, where 10\%, 50\%, and 100\% of the records were replicated respectively. For each of these datasets, we measured the baseline using the ML analysis.

\begin{figure}
\begin{center}
\includegraphics[width=0.85\linewidth]{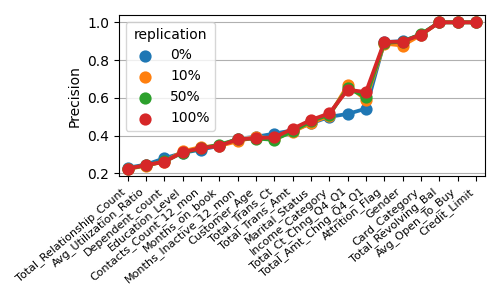}
\caption{Results comparing datasets with various amounts of record replication (none, 10\%, 50\%, and 100\%), for both categorical and continuous attributes. For most of the attributes, replication has no effect on the baseline precision measure. Two of the measures, however, show a modest effect.
}
\label{fig:replication}
\end{center}
\end{figure}

The results are shown in Figure~\ref{fig:replication}, which combines categorical and continuous attributes. Except for two attributes, the replication has little effect. The two attributes (both continuous) exhibit a relatively modest though certainly significant effect.

As expected, this provides evidence that avoiding overfitting is generally insensitive to dependent records, but is not necessarily foolproof. Other analysis techniques may of course be more sensitive to dependence.

\subsection{Comparison with prior work}

The core idea of comparing member and non-member datasets in order to assess vulnerability is not new.  The idea can be found in Fredrikson et al. (2014~\cite{ristenpart-pharmacogenetics}) and Yeom et al. (2018~\cite{yeom2018privacy}) studying the anonymity of ML models, in Stadler et al. (2020~\cite{stadler2020synthetic}) and Giomi et al. (2022~\cite{giomi2022unified}) studying the anonymity of synthetic data, in Kifer et al. studying differential privacy as applied to the 2020 US Census~\cite{kifer2022bayesian}, and in Kassem et al. more generally~\cite{kassem2019differential}.

\begin{figure}
\begin{center}
\includegraphics[width=1.0\linewidth]{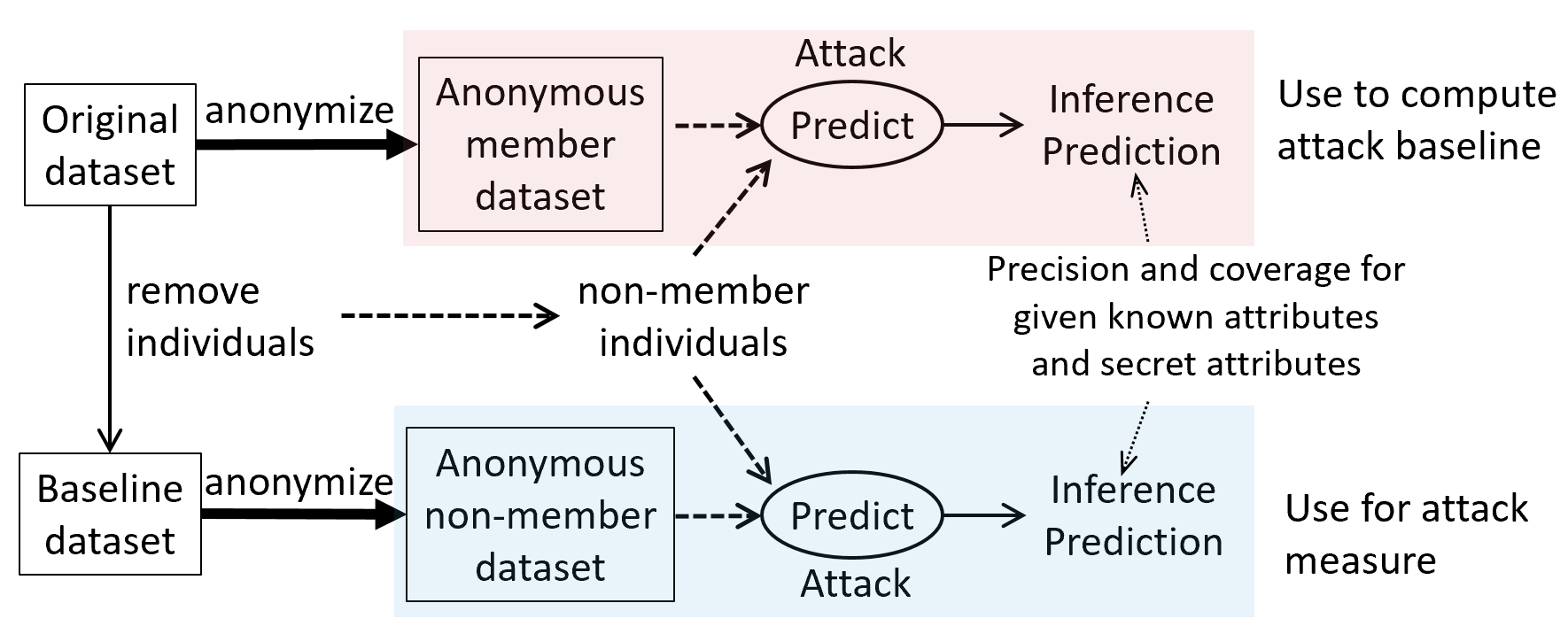}
\caption{Basic concept for prior work. The key difference is that prior work derives the baseline from the anonymous dataset rather than a non-anonymized dataset.
}
\label{fig:classic}
\end{center}
\end{figure}

There is a key difference in this prior work compared to the non-member framework. The purpose of the prior work is not to compute a baseline per se, but rather to measure the vulnerability of a given anonymization technology. Each of the prior works is some variation on the framework of Figure~\ref{fig:classic}. Instead of measuring predictions of a non-member against the original data, non-member predictions are made against an anonymized dataset.

By comparing non-member and member predictions against anonymized non-member and member datasets, the effectiveness of the anonymization method, relative to the given attack, is determined. An inference baseline per se is not computed.

Our non-member framework has two advantages over the prior work. First, it is more efficient to compute. The prior work requires an extra anonymization step.

\begin{figure}
\begin{center}
\includegraphics[width=0.85\linewidth]{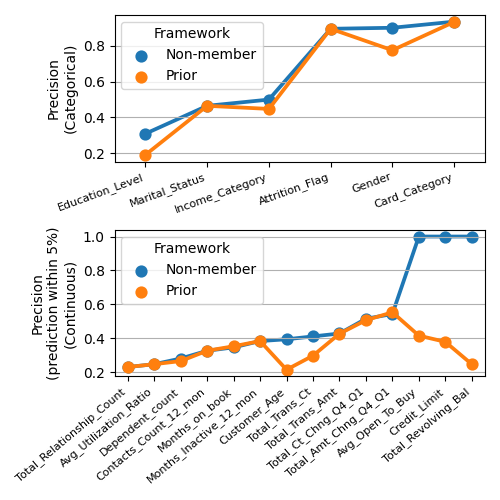}
\caption{Results comparing the non-member framework of Figure~\ref{fig:non-member} with the prior framework of Figure~\ref{fig:classic}. Both are using the same ML analysis.
}
\label{fig:nonVsPriorAcc}
\end{center}
\end{figure}

More importantly, our non-member framework produces a more accurate baseline. This is illustrated in Figure~\ref{fig:nonVsPriorAcc},  which compares the effect of the two frameworks. The precision measures for the non-member framework are the same as those in Figure~\ref{fig:allVsPiiAcc} where all attributes except the secret attribute are used as features.

To generate the precision measures for the prior framework, we generated a synthetic dataset from the same 7088 members using the open source CTGAN method provided by Synthetic Data Vault\footnote{https://sdv.dev/SDV/user\_guides/single\_table/ctgan.html}. The same ML model was run against the synthetic dataset.

From Figure~\ref{fig:nonVsPriorAcc}, we see that usually the prior framework produces a baseline identical or close to that of the non-member framework. For a few of the measures, however, the prior framework measure is much lower (e.g. \mytt{Total\_Revolving\_Bal}, roughly 0.2 compared to 1.0).

This creates two problems. First, it is possible that an attack substantially exceeds the artificially low baseline and so is misinterpreted as a vulnerability.

The bigger problem is that the prior framework makes it hard to assess if one is too-aggressively anonymizing. Take for example the case of \mytt{Total\_Revolving\_Bal}. Suppose we are using the non-member framework, and so the baseline inference is $P_{base}=1.0$. Suppose that an attack measure on some anonymization scheme produces $P_{atk}=0.2$. It would be clear from this that the anonymization might be too aggressive. A weaker anonymization method with better utility might produce a higher $P_{atk}$ which nevertheless still falls below $P_{base}$.

By contrast, suppose we are using the prior framework with $P_{base}=0.2$. An attack with $P_{atk}=0.2$ would tell us that the anonymization method was strong enough, but we would not know that we have room to improve utility.

In addition to the prior work cited above, note that the idea of comparing datasets that differ by a single individual can be used to measure differential privacy. There are many examples of this, but Gehrke et al. in particular is similar to our non-member framework in that it generates a baseline dataset from the (sampled) original data rather than from anonymized data~\cite{gehrke2011towards}. This work is used to provide a definition of differential privacy rather than generate a baseline per se. Other examples can be found in~\cite{desfontaines2020sok}.

Note finally that the new reconstruction attack on the 2010 US Census refers to the prior framework~\cite{abowd2023census}, but does not implement it on the grounds of its being too expensive.

\subsection{GDPR and the non-member framework}

We are interested in the question of whether the European Union data protection regulation GDPR supports the idea of an allowed baseline inference.
At its core, GDPR Article 4(1) defines personal data as:
\begin{noindent}
\begin{quoting}
\textit{any information relating to an identified or identifiable natural person (`data subject')}
\end{quoting}
\end{noindent}
Identifying includes by direct means (e.g. name, account number) and indirect means (e.g. location trace). While GDPR only applies to personal data (Article 2 (1), recital 26) defines anonymous as:
\begin{noindent}
\begin{quoting}
\textit{personal data rendered anonymous in such a manner that the data subject is not or no longer identifiable is not considered personal data.
}
\end{quoting}
\end{noindent}

The first question is, does GDPR support the core concept of the non-member framework; that if an individual is not present in a dataset, then the release of that dataset does not violate that individual's privacy (i.e. is not considered personal data with respect to that individual)? Since the very definition of personal data requires the presence of an `identified or identifiable natural person' in the data, we take it as self-evident that the lack of presence means that the person cannot be identified and therefore the data cannot violate the person's privacy.

The second question is, does GDPR support the concept of an allowed inference baseline? The GDPR itself intentionally says very little about how to determine if data is anonymous, and in particular says nothing about the role of inference in anonymity. The Article 29 Data Protection Working Party, however, in its ``Opinion 05/2014 on Anonymisation Techniques'', does address inference~\cite{article29}. It identifies inference as one of three criteria against which anonymization can be evaluated:

\begin{noindent}
\begin{quoting}
     \textit{Inference, which is the possibility to deduce, with significant probability, the value of an attribute from the values of a set of other attributes.}
\end{quoting}
\end{noindent}

A key phrase here is ``significant probability''. The Article 29 opinion elucidates what it means by ``significant probability'' through several examples. Primary among these is the example of k-anonymity and the Homogeneity Attack~\cite{machanavajjhala2007diversity}. This is an attack where, for a given set of known attribute values, there is only one possible value for the secret attribute. In this case, the probability of a correct inference is 100\%, which is indeed significant.

Nevertheless, the criteria does not say ``100\% probability'', so it implicitly recognizes a probability less than 100\% to potentially break anonymity. In discussing the Permutation mechanism (swapping values between records), the Article 29 opinion says this:

\begin{noindent}
\begin{quoting}
\textit{not knowing which attributes have been permutated, the attacker has to consider that his inference is based on a wrong hypothesis and therefore only probabilistic inference remains possible.}
\end{quoting}
\end{noindent}

In other words, the Article 29 opinion recognizes a ``probabilistic inference'' which does not break anonymity. Though the Article 29 opinion does not elucidate how to determine when an inference is merely probabilistic, versus when it breaks anonymity, this example suggests that GDPR supports the general idea of an allowed baseline inference.

It is important to recognize that the allowed inference baseline is a probabilistic measure: the precision as measured across a \textit{group of individuals}. The behavior relative to any single given individual is, however, not probabilistic.

Suppose there is an individual target $T$ where the inference prediction made by an analysis on the baseline dataset for $T$ (as a non-member) is incorrect, but an inference prediction for $T$ made by attack on the anonymous dataset (where $T$ is a member) generates a correct prediction. Seemingly this particular individual's privacy has been compromised, even if the statistical attack precision is below the baseline precision.

Here we rely on the uncertainty of the attacker. While the prediction for $T$ went from wrong to right in point of fact, the attacker doesn't know this. The uncertainty of the attacker protects the target. This leads to the third question: does the Article 29 opinion recognize uncertainty as a legitimate form of protection?

The answer is `yes'. For instance, when discussing inserting noise, the Article 29 opinion says:
\begin{noindent}
\begin{quoting}
\textit{even if the noisy disclosure mechanism is known in advance, the privacy of the data subject is preserved, since a degree of uncertainty remains.}
\end{quoting}
\end{noindent}

The allowed baseline essentially establishes a degree of uncertainty which by definition must be anonymous relative to a given individual, since that individual is not present in the dataset from which the baseline was derived. If the degree of uncertainty is the same or greater when the individuals are present in the anonymized dataset, then the individuals' privacy is equally or better protected. Since this is true for every individual in the anonymized dataset, the dataset may be regarded as anonymous by GDPR standards, and therefore non-personal data.

We believe that the non-member framework can make a contribution to the definition of anonymity in the GDPR and other privacy regulations. By defining an allowed baseline inference, it can be stated that any anonymization technique that only allows inferences below this baseline is certainly anonymous, at least with respect to the known inference attacks.


%
%
%
%

%% file: base-rate.tex
\section{Base rate errors}
\label{sec:base-rate}

In a membership attack, an adversary tries to determine if a target individual is a member of the original dataset. This attack gained notice in 2008, when Homer et al.~\cite{homer-genetics} demonstrated the ability to determine if a target, given knowledge of the target's DNA, is included in the summary statistics of a Genome-wide Association study (GWAS). Interest in membership attacks grew in 2017 when Shokri et al.~\cite{shokri2017membership} demonstrated the ability to determine if a target was included in the training set of a machine learning (ML) model.

Shokri et al. claimed that their results \textit{``have substantial practical privacy implications''}, and \textit{``indicate that membership inference can present a risk to health-care datasets.''} The 2021 survey paper by Hu et al.~\cite{hu2022membership} catalogues 65 ML membership attack papers, and makes a stronger claim that \textit{``membership inference attacks raise severe privacy risks to individuals''}. Given these statements, along with the sheer volume of work on these attacks, one would expect to see clear evidence that these attacks are indeed severe. In this section, we give examples of how these papers fall prey to the base rate error, and in so doing fail to provide this evidence.


The core problem with pretty much this entire body of work is that, with rare exception attack effectiveness is reported based on balanced observations: an equal number of members and non-members. In virtually any realistic scenario, however, the observations are skewed: there is far greater chance that an observation is a non-member. This increases false positives, making attacks less effective than presented, especially if precision and recall are measured.

Reporting on balanced observations is useful for determining the generic behavior of attacks independent of the actual deployment scenario, and for apples-to-apples comparisons between different attacks. This style of reporting is well suited to studying the attacks themselves but not to assessing the actual risk in real deployments.

The idea that CS researchers should be interested in studying vulnerabilities but not risk is perfectly reasonable. A division of labor where the CS researcher describe and measure the vulnerabilities, and then stakeholders use those vulnerabilities to assess the risk as it pertains to them, makes sense.

Unfortunately, the standard approach to measuring membership attacks does not make it easy to assess risk, and in most cases makes it impossible. As discussed in Section~\ref{sec:precision}, an intuitive measure for assessing risk is precision and coverage. Precision tells us the likelihood that a prediction is correct, and coverage tells us for what fraction of individuals we can get a given prediction. Either low precision or low recall makes an attack less attractive to an attacker, and therefore incurs less risk.

A variety of measures are used by different membership attack papers (\cite{hu2022membership} lists ROC curves, accuracy, precision, recall, advantage, F1 score, and AUC). Precision and recall over balanced observations yields an incorrect measure if the actual attack scenario is over skewed observations. ROC curves have become the most common measure in membership attack papers.  Fortunately, ROC curves (False Positive Rates $FPR$ against True Positive Rates $TPR$) can be translated into precision and recall for any observational skew using:

\ifccs
\vspace*{-1cm}
\else
\fi
\begin{align}
    TPR &= TP/M \\
    FPR &= FP/N \\
    recall &= TPR \\
    precision &= TP / TP+FP \\
    &= (TPR\!*\!M) / ((TPR\!*\!M)\! +\! (FPR\!*\!N))
    \label{eq:prec}
\end{align}

where $M$ is the number of members, $N$ is the number of non-members, $TP$ is the number of true positives, and $FP$ the number of false positives. $M$ and $N$ define the observational skew.

So while in principal precision and recall for any observational skew can be computed from an ROC curve, in practice the majority of membership attack papers plot ROC curves in such a way that high-precision/low-recall points cannot be determined.

\begin{figure*}
\begin{center}
\includegraphics[width=1.0\linewidth]{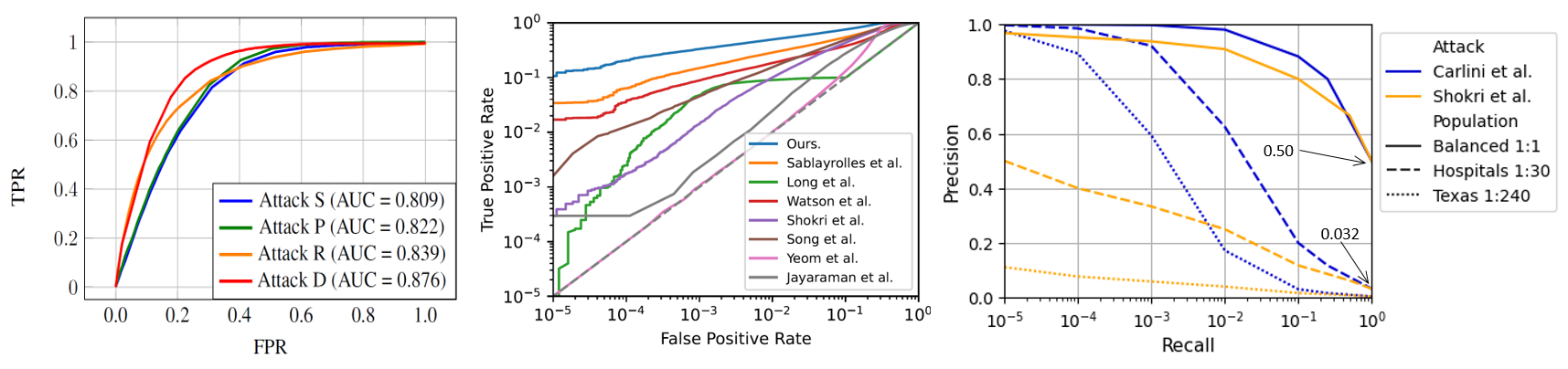}
\caption{Three examples of reporting membership attack effectiveness. The left-most plot is copied verbatim from Ye et al.~\cite{ye2022enhanced}. It cannot be determined if there is a high-precision/low-recall regime, since the plot lacks detail at low FPR. The middle plot, which has no relation to the leftmost plot, is copied verbatim from the arXiv version of Carlini et al.~\cite{carlini2022membership}. By using log-log scale, Carlini et al. exposes detail at low FPR. This allows us to generate the third plot, which maps two of the ROC curves of Carlini et al. to precision-recall curves for different observation skews.
The curve labeled ``Ours'' in the middle plot corresponds to ``Carlini et al.'' in the rightmost plot. The observational skews for the three populations are shown as \emph{member:non-member}. 
}
\label{fig:roc-pr-curves}
\end{center}
\end{figure*}

Consider for instance the left-most plot of Figure~\ref{fig:roc-pr-curves}. This is a set of ROC curves taken verbatim from Ye et al.~\cite{ye2022enhanced}. The y axis is $TPR = recall$. The x axis $FPR$ can be thought of as the inverse of precision: lower values equate to higher precision. Most of the curve lies in the low precision regime, and so indicates low risk. The high precision part of the curve is compressed into the tiny space at the lower left of the plot, and therefore does not provide enough detail to evaluate precision and recall in the higher precision regime where there may be risk.

Carlini et al. recognizes this problem~\cite{carlini2022membership}, and suggests measuring attacks at low FPR, and then plotting the ROC curves on a log-log scale to adequately capture the high precision part of the curve. Rezaei and Liu have also suggested that false positives are generally under-reported~\cite{rezaei2021difficulty}. The middle plot of Figure~\ref{fig:roc-pr-curves} is taken verbatim from the arXiv version of Carlini et al.~\cite{carlini2022membership}, exposing detail not discernable in the leftmost non-log plot (noting that the two plots are using different data and are therefore not related per se).

The log-log approach of Carlini et al. does allow one to derive meaningful precision-recall curves from the ROC curves using equation~\ref{eq:prec}. We did exactly this by taking a sample of points from the ROC curves of the middle plot (listed in Figure~\ref{fig:carlini_points}), applying equation~\ref{eq:prec} using realistic observational skews, and replotting them as precision-recall curves in the rightmost plot.  This allows us to mimic the risk analysis that a stakeholder working with this data might undertake.

\begin{figure}
\begin{center}
\begin{small}
\begin{tabular}{lll}
    \toprule
    FPR & Shokri TPR & Carlini TPR \\
    \midrule
    0.00001 & 0.0003 & 0.1 \\
    0.0001 & 0.002 & 0.2 \\
    0.001 & 0.015 & 0.35 \\
    0.01 & 0.1 & 0.5 \\
    0.1 & 0.4 & 0.75 \\
    0.25 & --- & 1.0 \\
    0.5 & 1.0 & --- \\
    1.0 & 1.0 & 1.0 \\
     \bottomrule
 \end{tabular}
 \end{small}
 \caption{FPR and TPR values as read from the middle plot of Figure~\ref{fig:roc-pr-curves} to produce the right plot.}
 \label{fig:carlini_points}
 \end{center}
 \end{figure}

These two plots use the dataset that Shokri et al.~\cite{shokri2017membership} derived from the Texas Hospital Discharge Data Public Use Data\footnote{\href{https://www.dshs.texas.gov/THCIC/Hospitals/Download.shtm}{https://www.dshs.texas.gov/THCIC/Hospitals/Download.shtm}}.
The middle plot of Figure~\ref{fig:roc-pr-curves} shows the results of different membership attacks on the Texas hospital dataset (Figure 20 from the arXiv version of ~\cite{carlini2022membership}). The rightmost plot converts two of the ROC curves to their corresponding precision-recall curves for three different skews\footnote{We chose Carlini et al. because it is the most effective attack, and Shokri et al. simply because it is the pioneering work.}.

Shokri et al. describe their dataset as coming from ``several health facilities'', and contains a total of 67K records. Since Shokri et al. did not select records based on any criteria (i.e. having had a certain disease), the only meaningful membership inference an attacker could make on an ML model generated from this data is to predict whether the target has visited a hospital or not.

We simulate two scenarios:
\begin{itemize}
    \item The attacker knows which hospitals are included in the study, and that if the target visited any hospital, it is one of these with high probability.
    \item The attacker only knows that all records are from Texas.
\end{itemize}

As a rough but reasonable estimate, we assume that 20\% of individuals have visited a hospital in one of the four years\footnote{\href{https://www.statista.com/statistics/184447/us-population-with-a-hospitalization-by-age/}{https://www.statista.com/statistics/184447/us-population-with-a-hospitalization-by-age/} states that roughly 7\% of the US population is hospitalized in any given year, but many of these will be the same individuals.}.

For the first scenario, there is a 1/5 chance that the target visited a hospital and is therefore in the dataset. Shokri et al. used a training set of 10K records, which is roughly 1/6 of the dataset. Therefore the skew for the first scenario is 1:30.  For the second scenario, the population of Texas is 2.4M, 10K of whom are in the training set. The skew is therefore 1:240.  We also include a curve for the balanced scenario 1:1, even though this scenario is literally impossible given the 1:6 sampling.

Compared to the ROC curve, the precision-recall curves allow a stakeholder to directly reason about risk without having to go through the exercise of translating ROC into precision-recall.

As expected, skew has a marked effect on attack performance. It is not our place to decide if a given precision-recall represents excessive risk, but looking at the Hospitals curves, which represent the best-case scenario for the attacker, this strikes us as extremely low risk. Shokri et al. manage only 40\% precision for 1/10000 targets. We simply can't see how this justifies the statement that membership inference attacks present a risk to health-care datasets.

Even assuming the much stronger Carlini et al. attack, an attacker can get 95\% precision for only 1 in 8000 targets.  Even at 80\% precision, where the target in any event has substantial deniability, the recall is only 1 in 800.

Finally, note that neither of these attacks have better precision than simply predicting that every individual is a member (recall of 1.0). Though out of scope for this paper, it would interesting to explore whether these membership inference attacks exceed an allowed statistical baseline for other recall values.

We recommend that membership inference attack papers report attack effectiveness as a series of precision-recall curves using a common set of observational skews, for instance 1:2, 1:5, 1:10, 1:50 etc. Doing so would facilitate risk assessments while still allowing researchers to make apples-to-apples comparisons of attack effectiveness.

\subsection{Other base rate errors}

This failure to account for observational skew is not limited to GWAS and ML applications. The several studies that look at membership in genomic data-sharing beacons used balanced observations~\cite{shringarpure2015privacy}\cite{raisaro2017addressing}\cite{von2019re}. Balanced observations are used by Pyrgelis et al.~\cite{pyrgelis2017knock} looking at membership in location datasets, and Stadler et al.~\cite{stadler2020synthetic} looking at membership in synthetic data. As part of a bounty program on anonymity~\cite{diffix-bounty-2020-tr}, Francis measures membership attack effectiveness assuming balanced observations (thus oddly rewarding attackers for attacks that are almost certainly not effective in practice).

Finally, note that others have criticized the implicit ``base rate fallacy'' associated with the GWAS membership attack papers, including Braun et al. in 2009~\cite{braun2009needles} and Erlich et al. in 2014~\cite{erlich2014routes}. The point is also made in the 2016 NIH National Human Genome Research Institute (NHGRI) workshop that concluded that the risks of this attack are small, and do not justify continued use of restrictions placed on GWAS summary statistics~\cite{nhgri-study}.

%% file: discussion.tex
\section{Discussion}
\label{sec:discussion}

It is important to note that none of the attack papers examined in this paper make false statements per se. Each paper proposes an attack measure (reconstruction percentage, ROC over balanced observations, etc.), and presents results according to that measure. Each paper expresses an opinion that, according to its measure, there is a privacy risk worth consideration. In other words, each paper defines a vulnerability bar above which one should be cautious.

What this paper is essentially proposing is that that vulnerability bar can in many cases be safely lowered. In discussing this idea with colleagues, we have heard two objections. First, that caution is a good thing; that it is better to be safe than sorry. Second, that if attack papers follow our measurement approach, they will appear less risky and therefore are less likely to be published, leading to fewer researchers exploring attacks on anonymized data. We discuss these objections in turn.

\subsection{The ``caution is good'' argument}

If the bar is set too high, then data that might otherwise get released is held back or too aggressively distorted. This harms the generation of new knowledge. If the bar is set too low, then unsafe data may be attacked by bad actors leading to real privacy violations. While there is evidence of the former, there no evidence of the latter.

Regarding the former, the first US Census Bureau reconstruction attack resulted in the Bureau redesigning its methodology, leading to loss of analytic utility~\cite{kenny2021use, santos2020differential,hauer2021differential,santos2021changes, santos2020differential,mueller20222020,winkler2021differential}, delays in the release of data, and ultimately two lawsuits contesting the need for the redesign~\cite{lawsuit-alabama,lawsuit-fairlines}. One could argue that the new reconstruction attack retroactively validates the decision to redesign, but it is also possible that a less disruptive solution could have been found.

The Facebook release of election data for the Social Science Research Council~\cite{ssrc-2019} was so heavily anonymized that one research group deemed the data ``nearly useless''~\cite{hegelich2020facebook}. The ability of non-Facebook researchers to study the impact of Facebook usage on elections was substantially weakened.

In response to the publication of a membership attack on Genome-wide Association (GWA) data~\cite{homer-genetics}, the NIH National Human Genome Research Institute (NHGRI) imposed access controls on GWAS summary statistics, thus effectively reducing usage of the data~\cite{nhgri-study}. (These controls have since been removed.)

Regarding the latter, we have made a concerted effort to find reports of malicious reidentifications, and cannot find any. In addition to conducting online searches, we have asked dozens of colleagues, directly or via professional privacy mailing lists\footnote{UK Anonymisation Network (UKAN), Internet Privacy Engineering Network (IPEN)}, for examples of malicious attacks. This includes members of several national data protection authorities in Europe, and members of the privacy teams of two national census bureaus. We have asked individuals involved in the release of medical data, including a manager at the AHRQ HCUP databases~\cite{hcup}.  We have asked the curator of databreaches.net.  We made a small effort to find evidence of reidentified data for sale on the dark web (e.g. DarkFox Market), but it is not immediately clear how to search for reidentified data per se. Finally, we can find no mention of government surveillance programs, as revealed by Wikileaks and the Snowden documents, that aim to reidentify anonymized data.

As of 2018, the NIH was unaware of any reports of attacks on genomic summary results~\cite{nih2018} despite papers warning of privacy vulnerabilities\cite{homer-genetics}\cite{wang-identity-genome} and multiple years of summary results being openly published.

It is worth noting that we cannot even find reports of malicious exploitation of the New York City taxi data~\cite{nyc-taxi}, which yearly publicly releases a pseudonymized database of all taxi rides including high-precision times and locations. Indeed we cannot find malicious attacks on any pseudonymized data release. We are not at all suggesting here that pseudonymized data should be considered non-personal. Rather we are simply pointing out that even weak methods appear to be pretty effective in practice.

Of course lack of evidence does not mean that malicious attacks never happen. Indeed, we would expect such evidence to be harder to find since, unlike most data breaches, an attack on released data leaves no online trace. On the other hand, if attackers are to exploit what they have learned, they would normally need to reveal to someone what they have learned, for instance by publishing the embarrassing information or by selling it online. This in turn could lead to exposing how the information was learned. It seems likely that if malicious attacks on anonymized data were happening on a large scale, there would be some evidence of it.

In balance, we believe that lowering the vulnerability bar by adapting the suggestions of this paper (the allowed inference baseline and precision/coverage metrics on skewed observations) would be overall beneficial.

\subsection{The ``less research on attacks'' argument}

It is very important that researchers find vulnerabilities before bad actors do. In a perfect world, we would deploy formal private anonymization methods and not worry about vulnerabilities, but utility usually trumps provability and so most anonymization is informal. It must therefore remain the case that researchers are more incentivized than bad actors.

If program committees require that attack papers use precision and coverage measures, and take into account baseline inferences and realistic base rates, then vulnerabilities will appear less severe than they currently tend to do. It would be a bad thing if PCs then starting rejecting papers they would accept today.

We recommend that, so long as an attack is novel and interesting, PCs should \textbf{NOT} downgrade or reject papers simply because the attack does not appear high risk.  After all, today's low-risk but novel attack may inspire tomorrow's high-risk attack. In other words, the primary consideration in accepting an attack paper should be its novelty and academic interest, not its risk severity.

%% file: summary.tex
\section{Future Work}
\label{sec:summary}


This paper only provides a conceptual framework, algorithm sketch, and simple demonstration of the non-member framework. Considerable future work is required to flesh out the idea, understand and fix flaws, and build tools that can produce allowed inference baselines on any dataset.

While we have found a few examples of allowed inference errors, we have not done a thorough examination of the literature. We do not know how common the error is, and encourage researchers to look for more instances. We also encourage researchers to explore whether other types of measurement errors exist in the anonymity attack literature.

Finally, we hope that this paper leads to more accurate and uniform reporting of data anonymity vulnerabilities.